\documentclass[a4paper,12pt]{article}
\usepackage[russian,english]{babel}
\usepackage {amssymb,amsmath,latexsym,enumerate,graphicx}
\usepackage{indentfirst}
\usepackage{titlesec}
\usepackage {subfigure,epsfig}
\usepackage{caption2}
\usepackage[cp1251]{inputenc}
\usepackage{cite}

\numberwithin{equation}{section}

\makeatletter

\renewcommand{\@biblabel}[1]{#1.} \makeatother

\makeatletter
\renewcommand{\section}{\@startsection%
{section}%
{1}%
{17pt}%
{-3.5ex plus -1ex minus -.2ex}%
{-10pt}%
{\textbf}%
} \makeatother

\begin{document}
\null
  \vskip 1.5em%
  \begin{center}%
        {\begin{tabular}[t]{c}%
        \large \lineskip .5em \textbf{Nikolai A. Kudryashov and Olga Yu. Efimova}
        \end{tabular}\par%
        }\vskip 1.5em%
      {\textbf{POWER EXPANSIONS FOR SOLUTION OF
     THE FOURTH-ORDER
      ANALOG  TO THE FIRST PAINLEV\'{E} EQUATION} \par}%
     \end{center}%
 \par
  \vskip 1.5em
\begin{abstract}
\end{abstract}
{\footnotesize
\begin{quotation} One of the fourth-order analog to the first Painlev\'{e} equation is studied.
All power expansions for solutions of this equation near points
$z=0$ and $z=\infty$ are found by means of the  power geometry
method. The exponential additions to the expansion of solution near
$z=\infty$ are computed. The obtained results confirm the hypothesis
that the fourth-order analog of the first Painlev\'{e} equation
determines new transcendental functions.
\end{quotation}
}
\bigskip

\section{\!\!\!\!\!\!.\,\, Introduction.}

More than one century ago Painlev\'{e} and his school discovered six
irreducible second-order equations, which determine new
transcendental functions. Over a long period of time these functions
seemed to have no physical applications, but now these equations are
widely used for description of different physical processes. At the
present the problem of analysis of the higher analogs to the
Painlev\'{e} equations has appeared. There is a lot of works,
devoted to the solution of this problem
\cite{Kudryashov01,Kudryashov02,Kudryashov0/5,Kudryashov06,Kudryashov07,
Clarkson01, Cosgrove01, Joshi01, Pickering01, Hone, Kawai01,
Mugan01, Mugan02}. One of the fourth-order analogs of the first
Painlev\'{e} equation is equation
\begin{equation}
\label{1.5} w_{zzzz} +18w\,w_{zz}+9w^2_z+24w^3=z
\end{equation}

In papers \cite{Kudryashov0/5, Kudryashov0/6, Kudryashov02,
Kudryashov05, Kudryashov10, Pickering01} it was shown that equation
\eqref{1.5} has properties, that are typical for  the Painlev\'{e}
equations $P_1\div P_6$. Equation \eqref{1.5} belongs to the class
of exactly solvable equations, as it has Lax pair and a lot of other
typical properties of the exactly solvable equations. However it
does not have the first integrals in the polynomial form, that is
one of the features of  the Painlev\'{e} equations. Equation
\eqref{1.5} seems to determine new transcendental functions just as
equations $P_1\div P_6$, although the rigorous proof of the
irreducibility of equation \eqref{1.5} is now the open problem.

Thereupon the study of all the asymptotic forms and power expansions of equation
\eqref{1.5} is the important stage of the analysis of this equation, as this fact
indirectly confirms the irreducibility of equation \eqref{1.5}.

This equation does not have the exact solutions, and so it is very
important to find the asymptotic forms and the power expansions of
the solution of this equation, that is the aim of this work.

Let us find all the power expansions for the solution of equation \eqref{1.5} in the
form of

\begin{equation}
\label{1.5.a} w(z)= c_r\,z^{r}+ \sum_{s}\,c_s\,z^{s}
\end{equation}
at $z\rightarrow\,0$, then $\omega=-1$, $s>r$ and at $z\rightarrow\,\infty$, then
$\omega=1$, $s<r$.

For that we use the  power geometry method \cite{Bruno01,
Bruno02,Bruno03, Bruno-Kudryashov}.

The outline of this paper is as follows. Section 2 is devoted to the general properties
of equation \eqref{1.5}. In sections 3--5 the expansions near $z=0$ are found. In
sections 6--9 the power expansion near $z=\infty$ and its exponential additions are
obtained.

\section{\!\!\!\!\!\!.\,\,The general properties of equation \eqref{1.5}.}
Let us consider the fourth-order equation \eqref{1.5}

\begin{equation}
\label{1.7}f(z,w)\stackrel{def}{=}w_{zzzz} + 18ww_{zz} +9 w^2_z + 24 w^3  - z =0
\end{equation}

For monomials of equation \eqref{1.7} we have points $M_1=(-4,1),\,\,\,
M_2=(-2,2),\,\,\,M_3=(-2,2),\,\,\, M_4=(0,3),\,\,\, M_5=(1,0)$.

The carrier of equation is defined by four points $Q_1=M_1$,
 $Q_2=M_4 $,  $Q_3=M_5$  and  $Q_4=M_2=M_3$.  Their convex hull
$\Gamma$ is the triangle (fig. 1).

\begin{figure}[h] 
 \centerline{\epsfig{file=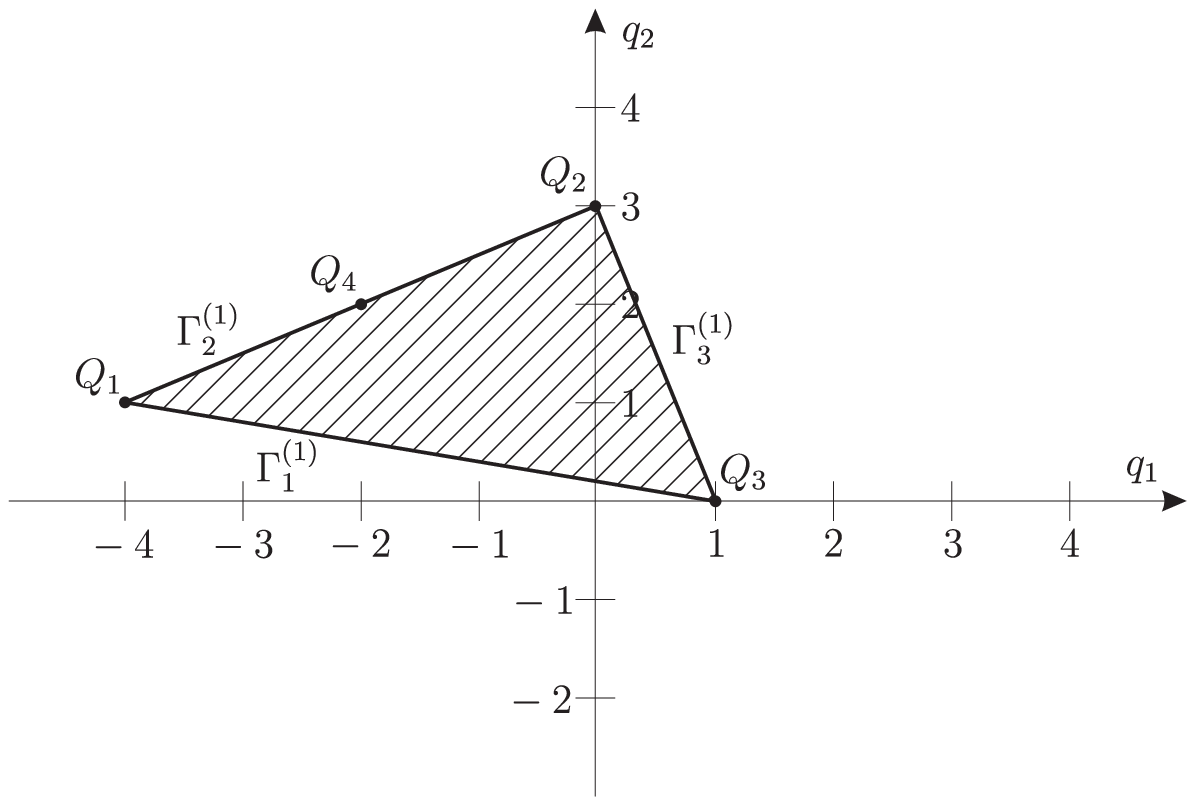,width=120mm}}
 \caption{}
\end{figure}

This triangle has apexes $Q_j\,(j=1,2,3)$ and edges $\Gamma_1^{(1)}=[Q_3,\,Q_1],\,\,
\Gamma_2^{(1)} =[Q_1,\, Q_2],\,\, \Gamma_3^{(1)} =[Q_2,\, Q_3]$

Outward normal vectors $N_j\,(j=1,2,3)$ of edges $\Gamma_j^{(1)} \, (j=1,2,3) $  are
determined by vectors

\begin{equation}
\label{1.8}N_1=(-1,-5),\,\,\, N_2=(-1,2),\,\,\, N_3=(3,1)
\end{equation}

The normal cones $U_j^{(1)}$ to edges $\Gamma_j^{(1)}$ are

\begin{equation}
\label{1.9}U_j^{(1)} =\mu N_j,\,\,\, \mu>0,\,\,\, j=1,2,3
\end{equation}

They and the normal cones $U_j^{(0)}$ of apexes $\Gamma_j^{(0)}=Q_j\,\, (j=1,\,2,\,3)$
are represented at fig. 2.

\begin{figure}[h] 
 \centerline{\epsfig{file=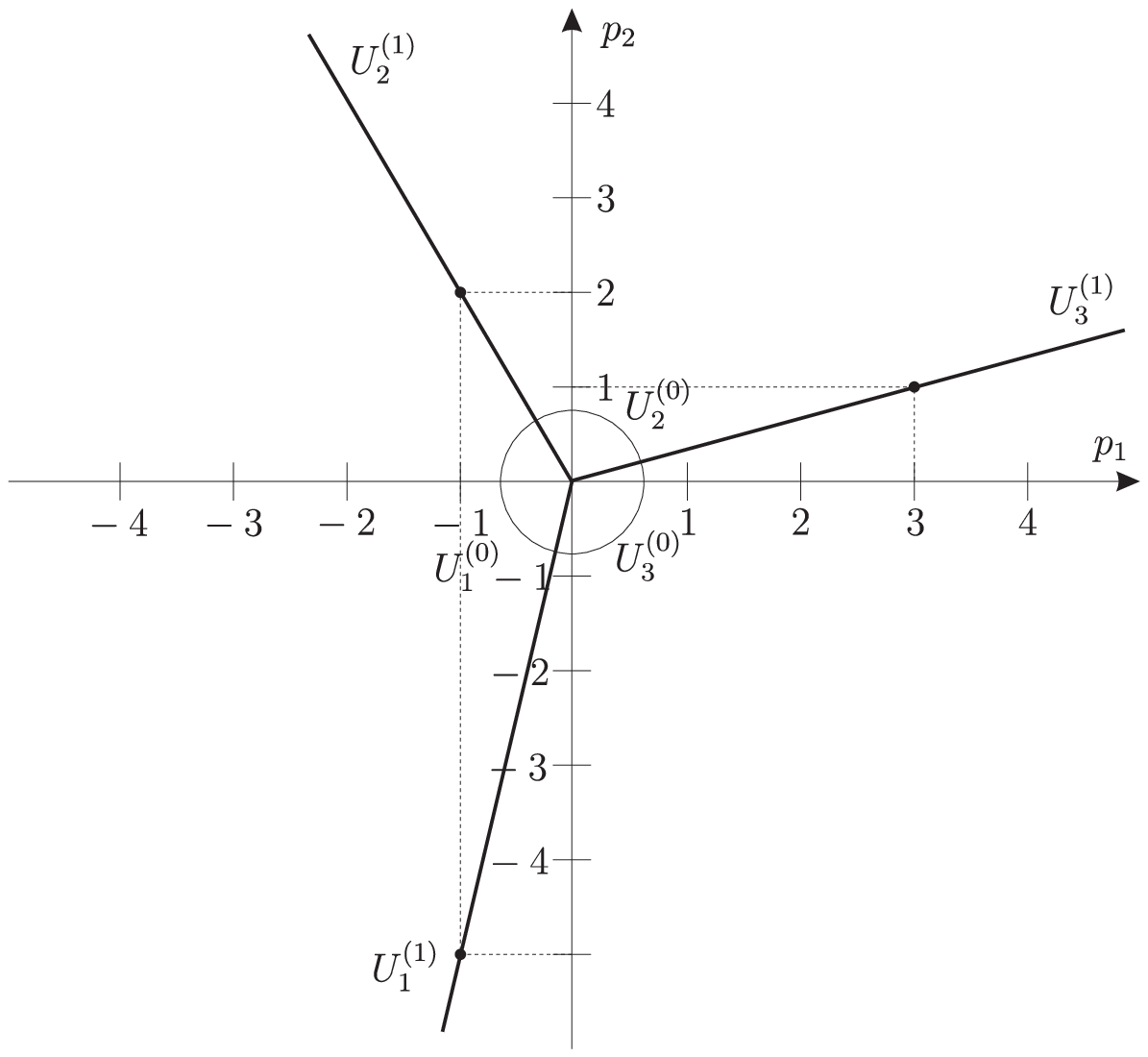,width=120mm}}
 \caption{}
\end{figure}

%

We can choose the basis of the lattice of the carrier of  equation \eqref{1.7} as

\begin{equation}
\label{1.9b}B_1=(-5,\,1),\,\,\quad\,B_2=(-3,\,2)
\end{equation}

Let us study solutions, corresponding to the bounds $\Gamma_j^{(d)},\,\, d=0,1;\,\,
j=1,\,2,\,3$ in view of the reduced equations, conforming to apexes $\Gamma_j^{(0)}
(j=1,\,2,\,3)$
\begin{equation}
\label{1.10}\hat{\emph{f}}_{1}^{(0)}\stackrel{def}{=}w_{zzzz}=0
\end{equation}
\begin{equation}
\label{1.11}\hat{\emph{f}}_{2}^{(0)}\stackrel{def}{=}24\,w^3=0
\end{equation}
\begin{equation}
\label{1.12}\hat{\emph{f}}_{3}^{(0)}\stackrel{def}{=}\,- z=0
\end{equation}
and reduced equations, conforming to edges $\Gamma_j^{(1)} (j=1,2,3)$
\begin{equation}
\label{1.13}\hat{\emph{f}}_{1}^{(1)}\stackrel{def}{=}w_{zzzz}- z=0
\end{equation}
\begin{equation}
\label{1.14}\hat{\emph{f}}_{2}^{(1)}\stackrel{def}{=}w_{zzzz} +18\,w\,w_{zz} +9w_z^2
+24 w^3=0
\end{equation}
\begin{equation}
\label{1.15}\hat{\emph{f}}_{3}^{(1)}\stackrel{def}{=}24w^3- z=0
\end{equation}

Note, that the reduced equations \eqref{1.11} and \eqref{1.12} are
the algebraic ones. According to \cite{Bruno02} they do not have
non-trivial power or non-power solutions.

\section{\!\!\!\!\!\!.\,\, Solutions, corresponding to apex $Q_1$.} Apex
$Q_1=(-4,1)$ is corresponded to reduced equation \eqref{1.10}.

Let us find the reduced solutions
\begin{equation}
\label{1.17}w=c_r z^r,\,\,\, c_r\neq0
\end{equation}
for $\omega (1,r) \in U_1^{(0)}$.

Since $p_1 < 0$ in the cone $U_1^{(0)}$, then $\omega=-1, \,\,\, z\rightarrow 0$ and
the expansions are the ascending power series of $z$. The dimension of the bound $d=0$,
therefor
\begin{equation}
\label{1.18}g(z,w)=w^4 \,w^{-1} \,w_{zzzz}
\end{equation}

We get the characteristic polynomial
\begin{equation}
\label{1.19}\chi(r)\stackrel{def}{=} g(z,z^r) = r(r-1)(r-2)(r-3)
\end{equation}
Its roots are
\begin{equation}
\label{1.20}r_1=0,\,\,\, r_2=1,\,\,\, r_3=2,\,\,\, r_4=3
\end{equation}

Let us explore all these roots.

The root $r_1=0$ is corresponded to vector $R=(1,0)$ and vector $\omega R\in
U_1^{(0)}$.

We obtain the family $\mathcal{F}_1^{(1)}\,1$ of reduced solutions $w=c_0$, where $c_0
\neq 0$ is arbitrary constant and $\omega = -1$. The first variation of equation
\eqref{1.10}
\begin{equation}
\label{1.21}\frac{\delta\hat{\emph{f}}_1^{(0)}}{\delta w} = \frac{d^4}{dz^4}
\end{equation}
gives operator
\begin{equation}
\label{1.22}\mathcal{L}(z)=\frac{d^4}{dz^4} \neq 0
\end{equation}
Its characteristic polynomial is
\begin{equation}
\label{1.23}\nu(k) =z^{4-k} \mathcal{L}(z)\, z^k =k(k-1)(k-2)(k-3)
\end{equation}
Equation
\begin{equation}
\label{1.24}\nu(k)=0
\end{equation}
has four roots
\begin{equation}
\label{1.25}k_1=0,\,\,\, k_2=1,\,\,\, k_3=2,\,\,\, k_4=3
\end{equation}

As long as $\omega=-1$ and $r=0$, then the cone of the problem is
\begin{equation}
\label{1.26}\mathcal{K}=\{k>0\}
\end{equation}
It contains the critical numbers $k_2=1,\,\,\, k_3=2$ and $k_3=3$. Expansions for  the
solutions, corresponding to reduced solution \eqref{1.17} can be presented in the form
\begin{equation}
\label{1.27}w=c_0 + c_1 z + c_2z^2 + c_3z^3 + \sum^{\infty}_{k=4} c_k z^k
\end{equation}
where all the coefficients are constants, $c_0\neq 0,\,\,\, c_1,\,\,\,c_2$, $c_3$ are
arbitrary ones and  $c_k \,\, (k\geq 4)$ are uniquely defined. Denote this family as
$\emph{G}_1^{(0)}1$. Expansion \eqref{1.27} with taking into account eight terms is
\begin{equation*}\begin{gathered}
\label{eq1.28} w \left( z \right) =c_{{0}}+c_{{1}}z+c_{{2}}{z}^{2}+c_{{3}}{z}^{3}-
 \left( \frac{3}{2}\,c_{{0}}c_{{2}}+{c_{{0}}}^{3}+\frac{3}{8}\,{c_{{1}}}^{2} \right)
{z}^{4}+\\+ \left( {\frac {1}{120}}-{\frac {9}{10}}\,c_{{0}}c_{{3}}-\frac{3}{5}\,
c_{{1}}c_{{2}}-\frac{3}{5}\,{c_{{0}}}^{2}c_{{1}} \right) {z}^{5}+\\+ \left(
\frac{1}{40} \,c_{{0}}{c_{{1}}}^{2}-\frac{1}{5}\,{c_{{2}}}^{2}-{\frac
{9}{20}}\,c_{{1}}c_{{ 3}}+{\frac
{7}{10}}\,{c_{{0}}}^{2}c_{{2}}+\frac{3}{5}\,{c_{{0}}}^{4} \right) { z}^{6}+\\+ \left(
\frac{3}{10}\,{c_{{0}}}^{2}c_{{3}}-{\frac {1}{280}}\,c_{{0}}+\frac{3}{
5}\,c_{{0}}c_{{1}}c_{{2}}+\frac{3}{5}\,{c_{{0}}}^{3}c_{{1}}-\frac{3}{10}\,c_{{2}}c_{{3}
}+\frac{1}{10}\,{c_{{1}}}^{3} \right) {z}^{7}+ \ldots
\end{gathered}\end{equation*}

Let us explore root $r_2=1$. The cone of the problem is $\mathcal{K}=\{k>1\}$. It
contains the critical numbers $k_2=2,\,\,\, k_3=3$. The expansion of solution,
corresponding to the reduced solution
\begin{equation*}\mathcal{F}_1^{(1)}2: \,\,\, w= c_1\,z\end{equation*}
can be written as
\begin{equation}
\label{1.29}w(z)=c_1z + c_2z^2 + c_3z^3 + \sum^{\infty}_{k=4} c_kz^k
\end{equation}
where $c_1\neq 0,\,\,\, c_2$ and $c_3$ are the arbitrary constants. Denote this family
as $\emph{G}_1^{(0)}2$. The expansion of solutions \eqref{1.29} with taking into
account seven terms is
\begin{equation*}\begin{gathered}
\label{1.29a}w(z)=c_{{1}}z+c_{{2}}{z}^{2}+c_{{3}}{z}^{3}-\frac{3}{8}\,{c_{{1}}}^{2}{z}^{4}+
 \left( {\frac {1}{120}}-\frac{3}{5}\,c_{{1}}c_{{2}} \right) {z}^{5}-\\- \left(
\frac{1}{5}\,{c_{{2}}}^{2}+{\frac {9}{20}}\,c_{{1}}c_{{3}} \right) {z}^{6}+
 \left( \frac{1}{10}\,{c_{{1}}}^{3} -\frac{3}{10}\,c_{{2}}c_{{3}}\right) {z}^{7}+\ldots
\end{gathered}\end{equation*}

For root $r_2=2$ the cone of the problem is $\mathcal{K}=\{k>2\}$. The critical number
is $k_3=3$. The expansion of the solutions, corresponding to the reduced solution
\begin{equation*}\mathcal{F}_1^{(1)}3: \,\,\, w= c_2\,z^{2}\end{equation*}
takes the form
\begin{equation}
\label{1.30}w=c_2z^2 +c_3z^3 + \sum^{\infty}_{k=4} c_k\,z^k
\end{equation}
Denote this family as $\emph{G}_1^{(0)}3$. Expansion \eqref{1.30} with taking into
account eight terms is
\begin{equation}\begin{gathered}
\label{eq1.30a}w(z)=c_{{2}}{z}^{2}+c_{{3}}{z}^{3}+{\frac
{1}{120}}\,{z}^{5}-\frac{1}{5}\,{c_{{2}}}
^{2}{z}^{6}-\frac{3}{10}\,c_{{2}}c_{{3}}{z}^{7}-{\frac {9}{80}}\,{c_{{3}}}^{2}
{z}^{8}-\\-{\frac {1}{630}}\,c_{{2}}{z}^{9}+ \left( {\frac {2}{75}}\,{c_{
{2}}}^{3}-{\frac {41}{33600}}\,c_{{3}} \right) {z}^{10}
 +\ldots
\end{gathered}\end{equation}

For root $r_3=3$ the cone of the problem is $\mathcal{K}=\{k>3\}$. There is no critical
number here. The expansion of solutions, corresponding to the reduced solution, is
\begin{equation*}\mathcal{F}_1^{(1)}4: \,\,\, w= c_3\,z^{3}\end{equation*}
takes the form
\begin{equation}
\label{1.31}w(z)=c_3z^3 + \sum^{\infty}_{k=4} c_k\,z^k
\end{equation}
Denote this family as $\emph{G}_1^{(0)}4$. The expansion \eqref{1.31} with taking into
account four terms is
\begin{equation}\begin{gathered}
\label{eq1.31a}w(z)=c_{{3}}{z}^{3}+{\frac {1}{120}}\,{z}^{5}-{\frac
{9}{80}}\,{c_{{3}}}^{2 }{z}^{8}-{\frac {41}{33600}}\,c_{{3}}{z}^{10}
 + \ldots
\end{gathered}\end{equation}

The expansions of solutions converge for sufficiently small $|z|$. The existence and
analyticity of expansions \eqref{1.27}, \eqref{1.29}, \eqref{1.30} and \eqref{1.31}
follow from Cauchy theorem.

\section{\!\!\!\!\!\!.\,\, Solutions, corresponding to edge $\Gamma^{(1)}_1$.} Edge $\Gamma^{(1)}_1$
is conformed by the reduced equation
\begin{equation}
\label{1.32}\hat{f}_1^{(1)} (z,y)\stackrel{def}{=}w_{zzzz} - z=0
\end{equation}
Normal cone is
\begin{equation}
\label{1.33}U^{(1)}_1 =\{-\mu(1,5),\,\,\mu>0\}
\end{equation}
Therefor $\omega=-1$, i.e. $z\rightarrow 0$ and $r=5$. Power solutions are found in the
form
\begin{equation*}
\label{1.34} w=c_5z^5
\end{equation*}
For $c_5$ we have
\begin{equation}
\label{1.35}c_5=\frac {1}{120}
\end{equation}
The only power solution is
\begin{equation}
\label{1.36}\mathcal{F}_2^{(1)}1: \,\,\,w=\frac{z^5}{120}
\end{equation}
Compute the critical numbers. The first variation of \eqref{1.13} is
\begin{equation}
\label{1.37}\frac{\delta \hat{f}^{(1)}_1}{\delta w} =\frac{d^4}{dz^4}
\end{equation}
We get the proper numbers
\begin{equation}
\label{1.38}k_1=0,\,\,\, k_2=1,\,\,\, k_3=2,\,\,\, k_4=3
\end{equation}
The cone of the problem
\begin{equation*}
\mathcal{K}=\{k>5\}
\end{equation*}
does not consist them.

Solution \eqref{1.36} is corresponded to two vector indexes $\tilde{Q}_1=(0,1),\,\,\,
\tilde{Q}_2= (5,0)$. There difference  $B=\tilde{Q}_1 - \tilde{Q}_2 =(-5,\,1)$ equals
to vector $Q_1-Q_2$. So solution \eqref{1.36} is conformed to lattice $\textbf{Z}$,
which consists of points $Q=(q_1,\, q_2)=k(-3,\,2) +m(-5,\,1)=(-3k-5l,\,2k+l)$, where
$k$  and $l$ are whole numbers. Points belong to line $q_2=-1$, if $l=-1-2k$. In this
case $q_1=5+7k$. As long as the cone of the problem here is $\mathcal{K}=\{k>5\}$, the
set of the carrier of solution expansion $\textbf{K}$ takes the form
\begin{equation}
\label{1.39}\textbf{K}=\{5+7n,\,\,\,n\in \mathbb{N}\}
\end{equation}
Then the expansion of solution can be written as
\begin{equation}
\label{1.40}w(z)=z^5\left(\frac{1}{120}+\sum^{\infty}_{m=1} c_{5+7m}\,z^{7m}\right)
\end{equation}
Expansion \eqref{1.40} with taking into account three terms is
\begin{equation}
\label{1.41}w(z)=\frac{z^5}{120}\left(1-{\frac {13}{31680}}\,{z}^{7}+{\frac
{601}{4911667200}}\,{z}^{14}+\ldots\right)
\end{equation}

Equation \eqref{1.32} does not have exponential additions and
non-power asymptotic forms.

\section{\!\!\!\!\!\!.\,\, Solutions, corresponding to edge $\Gamma^{(1)}_2$.}
Edge $\Gamma^{(1)}_2$ is corresponded to the reduced equation
\begin{equation}
\label{1.42}\hat{f}^{(1)}_2 (z,w) \stackrel{def}{=}w_{zzzz} -10\,w\,w_{zz}
-5\,w^2+10\,w^3=0\end{equation}

The normal cone is
\begin{equation}
\label{1.43}U^{(1)}_2=\{-\mu(1,-2),\,\mu>0\}
\end{equation}
Therefor $\omega=-1$, i.e. $z\rightarrow 0$ and $r=-2$. Hence the solution of equation
\eqref{1.42} we can find in the form
\begin{equation}
\label{1.44}w=c_{-2} z^{-2}
\end{equation}
For $c_{-2}$ we have the determining equation
\begin{equation}
\label{1.45}c_{-2}^2 +6\,c_{-2} +5 =0
\end{equation}
Consequently we get
\begin{equation}
\label{1.46}c_{-2}^{(1)} =-1,\,\,\quad\, c_{-2}^{(2)}=-5
\end{equation}
The reduced solutions are
\begin{equation}
\label{1.47}\mathcal{F}_2^{(1)}1: \,\,\, w=-z^{-2}
\end{equation}
\begin{equation}
\label{1.48}\mathcal{F}_2^{(1)}2: \,\,\, w=-5z^{-2}
\end{equation}
Let us compute the corresponding critical numbers. The first variation is
\begin{equation}
\label{1.49}\frac{\delta f_2^{(1)}}{\delta w}=\frac{d^4}{dz^4} +18 w_{zz}
+18w\frac{d^2}{dz^2} +18\,w_z \frac d{dz} + 72\,w^2
\end{equation}
Applied to  solution \eqref{1.47}, it produces operator
\begin{equation}
\label{1.50}\mathcal{L}^{(1)}(z) =\frac{d^4}{dz^4}-\frac{18}{z^2} \frac{d^2}{dz^2} +
\frac{36}{z^3} \frac{d}{dz}-\frac{36}{z^4}
\end{equation}
which is corresponded by the characteristic polynomial
\begin{equation}
\label{1.51}\nu(k ) =k^4 -6k^3 -7k^2 + 48k-36
\end{equation}
Equation
\begin{equation}
\label{1.52}\nu(k)=0
\end{equation}
has the roots
\begin{equation}
\label{1.53}k_1=-3,\,\,\, k_2=1,\,\,\, k_3=2,\,\,\, k_4=6
\end{equation}
With reference to solution \eqref{1.48} variation \eqref{1.49} gives operator
\begin{equation}
\label{1.54}\mathcal{L}^{(2)}(z) =\frac{d^4}{dz^4} - \frac{90}{z^2}
\frac{d^2}{dz^2}+\frac{180}{z^3} \frac{d}{dz} + \frac{1260}{z^4}
\end{equation}
which is corresponded by the characteristic polynomial
\begin{equation}
\label{1.55}\nu(k)=k^4 -6k^3 -79k^2 +264k +1260
\end{equation}
with roots
\begin{equation}
\label{1.56}k_1=-7,\,\,\, k_2=-3,\,\,\, k_3=6,\,\,\, k_4=10
\end{equation}
The cone of the problem here is
\begin{equation}
\label{1.57} \mathcal{K}=\{k>-2\}
\end{equation}
Therefor for the reduced solution \eqref{1.47} three critical numbers belong to the
cone, and there are two critical numbers for the reduced solution \eqref{1.48} in the
cone of the problem.

The set of the carriers of the solution expansions $\textbf{K}$ can be written as
\begin{equation}
\label{1.57a}\textbf{K}=\{-2+7n,\,\,\,n\in \mathbb{N}\}
\end{equation}
Sets $\textbf{K}(0)$, $\textbf{K}(0,3)$ and $\textbf{K}(0,3,6)$ are
\begin{equation} \begin{gathered}
\label{1.57b}\textbf{K}(1)=\{-2+7n+3m,\,\,n,m\in
\mathbb{N},\,\,n+m\geq\,0\}=\\
=\{-2,1,4,5,7,8,10,...\}
\end{gathered}\end{equation}
\begin{equation}\begin{gathered}
\label{1.57c1}\textbf{K}(1,2)=\{-2+7n+3m+4k,\,n,m,k\in
\mathbb{N},\,\,m+n+k\geq\,0\}=\\
=\{-2,1,2,4,5,6,7,8,...\}
\end{gathered} \end{equation}
\begin{equation}\begin{gathered}
\label{1.57c2}\textbf{K}(1,2,6)=\{-2+7n+3m+4k+8l,\,n,m,k,l \in \mathbb{N}
,\,\,m+n+k+l\geq\,0\}=\\
=\{-2,1,2,4,5,6,7,8,...\}
\end{gathered}\end{equation}
In this case the expansion for the solution of equation can be represented as
\begin{equation}\begin{gathered}
\label{eq1.57d}w(z)=\frac{2}{z^2}+ \sum^{}_{n+m+k+l>0}
c_{-2+7n+3m+4k+8l}\,z^{-2+7n+3m+4k+8l}
\end{gathered}\end{equation}
Denote this family as $G_{2}^{1}1$. The critical number $1$ does not
belong to set $\textbf{K}$, so the compatibility condition for $c_1$
holds automatically and  $c_1$ is the arbitrary constant. The
critical number $2$ also does not belong to sets $\textbf{K}$ and
$\textbf{K}(1)$, therefor the compatibility condition for $c_2$
holds too and $c_2$ is the arbitrary constant. But critical number
$6$ is a member of $\textbf{K}(1,2)$, so it is necessary  to verify
that the compatibility condition for $c_6$ holds and that $c_6$ is
the arbitrary constant. The calculation shows that in this situation
the condition holds and $c_6$ is the arbitrary constant too.
 The
three-parameter power expansion of solutions, corresponding to the reduced solution
\eqref{1.47} takes the form
\begin{equation}\begin{gathered}
\label{1.58}
w(z)=-\frac{1}{z^2}+c_{{1}}z+c_{{2}}{z}^{2}-\frac{3}{4}\,{c_{{1}}}^{2}{z}^{4}- \left(
\frac{3}{4}\,c_{{1}}c_{{2}}+{\frac {1}{96}} \right) {z}^{5}+c_{{6}}{z}^{6}+\\+{ \frac
{7}{25}}\,{c_{{1}}}^{3}{z}^{7}+{\frac {1}{4928}}\,c_{{1}}
 \left( 1848\,c_{{1}}c_{{2}}+17 \right) {z}^{8}+ \left( \frac{1}{8}\,{c_{{2}}}^{2}c_{{1}}-\frac{1}{4}\,c_{{1}}c
_{{6}}+{\frac {1}{448}}\,c_{{2}} \right) {z} ^{9}- \\-\left( {\frac
{1}{156}}\,{c_{{2}}}^{3}+{\frac {437}{5200}}\,{c_ {{1}}}^{4}+{\frac
{9}{52}}\,c_{{2}}c_{{6}} \right) {z}^{10} +\ldots
\end{gathered}
\end{equation}

The carrier of power expansion, corresponding to  reduced solution \eqref{1.48}, is
formed by the sets
\begin{equation}\begin{gathered}
\label{1.58a}\textbf{K}(6)=\{-2+7n+8m,\,n,m\in
\mathbb{N},\,\,m+n\geq\,0\}=\\
=\{-2,5,6,12,13,14,19,20,21,22,27,28,29,30,33,34,35,36,37,38,40,...\}
\end{gathered}\end{equation}
\begin{equation}\begin{gathered}
\label{1.58b}\textbf{K}(6,10)=\{-2+7n+8m+12k,\,n,m,k\in
\mathbb{N},\,\,m+n+k\geq\,0\}=\\
=\{-2,5,6,10,12,13,14,17,18,19,20,21,22,24...\}
\end{gathered}\end{equation}
The expansion for solution of equation can be written as
\begin{equation}\begin{gathered}
\label{eq1.58c}w(z)=-\frac{5}{z^2}+ \sum^{}_{n+m+k>0}
c_{-2+7n+8m+12k}\,z^{-2+7n+8m+12k}
\end{gathered}\end{equation}
Denote this family as $G_{2}^{1}2$. The critical numbers  6 and 10
do not belong to the set $\textbf{K}$ and the number 10 does not
belong to the set $\textbf{K}(6)$. For numbers 6 and 10 the
compatibility conditions holds automatically, therefor coefficients
$c_6$ and $c_10$ are the arbitrary constants. The two-parameter
expansion of solution, corresponding to the reduced solution
\eqref{1.48}, is
\begin{equation}\begin{gathered}
\label{1.59}w(z)=-\frac{5}{z^2}+{\frac
{1}{480}}\,{z}^{5}+c_{{6}}{z}^{6}+c_{{10}}{z}^{10} -{\frac
{1}{3502080}}\,{z}^{12}-\\-{\frac {1}{4480}}\,c_{{6}}{z}^{13}-{ \frac
{3}{68}}\,{c_{{6}}}^{2}{z}^{14}-{\frac {3}{24640}}\,c_{{10}}{z}^ {17} +\ldots
\end{gathered}\end{equation}
According to \cite{Bruno02}, the expansions of solutions
\eqref{1.58} and \eqref{1.59} do not have power and exponential
additions.

\section{\!\!\!\!\!\!.\,\, Solutions, corresponding to edge $\Gamma^{(1)}_3$.}
Edge $\Gamma^{(1)}_3$ is corresponded by the reduced equation
\begin{equation}
\label{1.60}\hat{f}_3^{(1)} (z,w) \stackrel{def}{=} 24\,w^3 - z=0
\end{equation}
In this case $\omega=1$, i.e. $z\rightarrow 0$ and $r=1/3$. The expansions are the
descending power series of $z$.

Reduced equation \eqref{1.60} has three power solutions
\begin{equation}
\label{1.61}\mathcal{F}_3^{(1)}1:\,\,\,\, w=\varphi^{(1)}(z)=
c_{1/3}^{(1)}\,z^{1/3},\,\quad\,c_{1/3}^{(1)}=\frac{1}{2}\sqrt[3]{\frac{1}{3}}
\end{equation}
\begin{equation}
\label{1.62}\mathcal{F}_3^{(1)}2:\,\,\,\, w=\varphi^{(2)}(z)=c_{1/3}^{(2)}\,
z^{1/3},\,\quad\,c_{1/3}^{(2)}\,=-\frac{1}{4}\left(1-
i\sqrt{3}\right)\sqrt[3]{\frac{1}{3}}
\end{equation}
\begin{equation}
\label{1.63}\mathcal{F}_3^{(1)}3:\,\,\,\,w=\varphi^{(3)}(z)=c_{1/3}^{(3)}\,z^{1/3},\,\quad\,c_{1/3}^{(3)}\,=-\frac{1}{4}\left(1+
i\sqrt{3}\right)\sqrt[3]{\frac{1}{3}}
\end{equation}
The shifted carrier of reduced solutions \eqref{1.61} -- \eqref{1.63} gives a vector
\begin{equation}
\label{1.64}B=\left(\frac13,-1\right)
\end{equation}
which equals a third of vector $Q_2-Q_1$. Therefor we explore the lattice, generated by
vectors $Q_3-Q_1$ and $B$. The basis of this lattice is $(-3,2)$ and $(1/3,-1)$. We
have $Q=(q_1,q_2) =k(-3,\,2) + m\left(\frac13,\,-1\right) =\left(-3k+ m/3,\,\,2k-m
\right)$, where $k$  and $m$ are the whole numbers. At the line $q_2=-1$ we have
$2k-m=-1$, wherefrom $m=2k+1$ and $q_1=\frac{(1-7k)}{3}$. And so the carrier of
solution is
\begin{equation}
\label{1.65}\mathbf{K}=\left\{k=\frac{1-7n}{3},\,\,\, n\in \mathbb{N}\right\}
\end{equation}
and the expansions of solutions take the form
\begin{equation}
\label{1.66}G_3^{(1)} l:\,\,\,\, w=\varphi^{(l)}(z)=c^{(l)}_{1/3} z^{1/3} +
\sum^{\infty}_{n=1} c^{(l)}_{(1-7n)/3}\, z^{{(1-7n)}/{3}}
\end{equation}
Here $c^{(l)}_{1/3}$ can be found from reduced solutions \eqref{1.61} -- \eqref{1.63},
coefficients $c^{(l)}_{(1-7n)/3}$ are computed sequentially. The calculating of the
coefficient $c_{-2}$ gives the result $c_{-2}=1/24$. The expansion of solution with
taking into account five terms is
\begin{equation}
\begin{gathered}
\label{1.66a}\varphi^{(l)}(z)=c_{{1/3}}\,{z^{1/3}}+\frac{1}{24}\,{z}^{-2}-{\frac
{1925}{46656}}\,{\frac {1}{c_{{ 1/3}}}}{z}^{-13/3}\,+\\
\\
+{\frac {509575}{3359232}}\,{\frac {1}{{c_{{1/3}}}^{2}}}\,{z}^{-{{20}/{3}}}-{\frac
{445712575}{362797056}}\,{\frac {1}{{c_{{1/3}}}^{3}}}{z}^ {-9}\,+ \ldots
\end{gathered}
\end{equation}
The obtained expansions seem to be divergent ones.

\section{\!\!\!\!\!\!.\,\, Exponential additions of the first level.}

Let us find the exponential additions to solutions \eqref{1.61}-\eqref{1.63}. We look
for the solutions in the form

\begin{equation*} \label{1.67}w=\varphi^{(l)}(z) + u^{(l)},\,\,\,
l=1,2,3
\end{equation*}
The reduced equation for the addition  $u^{(l)}$ is
\begin{equation}
\label{2.59}M_{l}^{(1)}(z) u^{(l)}=0
\end{equation}
where $M_{l}^{(1)}(z)$ is the first variation at the solution $w=\varphi^{(l)}(z)$. As
long as
\begin{equation}
\label{2.60}\frac{\delta f}{\delta w} =\frac{d^{\,4}}{dz^4} +18 w_{zz}
+18w\frac{{d\,}^2}{dz^2} +18w_z \frac{d}{dz} +72w^2
\end{equation}
then
\begin{equation}
\label{2.61}M_{l}^{(1)}(z) =\frac{d^{\,4}}{dz^4} +18\varphi^{(l)}_{zz} +18
\varphi^{(l)}\frac{d^{\,2}}{dz^2} +18\varphi_z^{(l)} \frac d{dz} + 72{\varphi^{(l)}}^2
\end{equation}
Equation \eqref{2.59} takes the form
\begin{equation}
\begin{gathered} \label{2.62}\frac{d^4u^{(l)}}{dz^4} +18 \varphi^{(l)}_{zz} u^{(l)}
+18 \varphi{(l)} \frac{d^2u^{(l)}}{dz^2} +18 \varphi^{(l)} _z \frac{du^{(l)}}{dz}
+72{\varphi^{(l)}}^2 u^{(l)} =0,\,\,\, \\
 l=1,2,3
\end{gathered}
\end{equation}
\begin{equation}
\label{2.63}\zeta^{(l)}=\frac{d \ln u^{(l)}}{dz}
\end{equation}
then from \eqref{2.63} we have
\begin{equation*} \label{}\frac{du^{(l)}}{dz} =\zeta^{(l)} u^{(l)},
\,\,\quad\, \frac{d^2u^{(l)}}{dz^2 }=\zeta _z^{(l)} u^{(l)} + {\zeta ^{(l)}}^2 u^{(l)}
\end{equation*}
\begin{equation*}
\label{}\frac{d^3u^{(l)}}{dz^3} =\zeta^{(l)}_{zz} u^{(l)}+ 3 \zeta ^{(l)} \zeta_z^{(l)}
u^{(l)} + {\zeta^{(l)}}^3 u^{(l)}
\end{equation*}
\begin{equation*}
\label{}\frac{d^4u^{(l)}}{dz^4} =\zeta^{(l)}_{zzz} u^{(l)}+ 4 \zeta
^{(l)}\zeta^{(l)}_{zz} u^{(l)} + 3 {\zeta_z^{(l)}}^2  u^{(l)} + 6 {\zeta^{(l)}}^2 \zeta
_z^{(l)} u^{(l)} + {\zeta ^{(l)}}^4 u^{(l)}
\end{equation*}
By substituting the derivatives
\begin{equation*}
\label{}\frac{du^{(l)}}{dz},\,\quad \, \frac{d^2 u^{(l)}}{dz^2} , \,\quad
\,\frac{d^4u^{(l)}}{dz^4}
\end{equation*}
into the equation \eqref{2.62} we get the reduced equation in the form
\begin{equation}
\begin{gathered}
\label{6.13}u^{(l)} \left[\zeta^{(l)}_{zzz} + 4\zeta^{(l)}\zeta^{(l)}_{zz}
+3{\zeta^{(l)}_z}^2 +6 {\zeta^{(l)}}^2 \zeta^{(l)}_z + \right. \\
+ \left. {\zeta^{(l)}}^4 +18\varphi_{zz}^{(l)} +18\varphi^{(l)} \zeta^{(l)}_z
+18\varphi^{(l)} {\zeta^{(l)}}^2 +18\varphi_z^{(l)} \zeta^{(l)}+72
{\varphi^{(l)}}^2\right]=0
\end{gathered}
\end{equation}

Let us find the power expansions for solutions of equation \eqref{6.13}. The carrier of
equation \eqref{6.13} consists of points
\begin{equation}
\begin{gathered}
\label{2.65}Q_1 =(-3,1),\,\,\, Q_2=(-2,2),\,\,\, Q_3=(-1,3), \\
Q_4=(0,4), \,\,\, Q_{5} =\left(\frac13,2\right),\,\,\, Q_{6} =\left(\frac23,0\right),
Q_{7}=\left(-\frac23,1\right),\\
Q_{8}=\left(-\frac53,0\right),\,\, Q_{5,k}=\left(\frac{1-7k}3,2\right),\,\,
Q_{6,k}=\left(\frac{2-7k}3,0\right),\\
Q_{7,k}=\left(-\frac{2+7k}3,1\right),\,\, Q_{8,k}=\left(-\frac{5+7k}3,0\right),\,\,\,k
\in \mathbb{N}
\end{gathered}
\end{equation}
The closing of convex hull of points of the carrier of equation \eqref{6.13} is the
strip. It is represented at fig. 3.

\begin{figure}[h] 
 \centerline{\epsfig{file=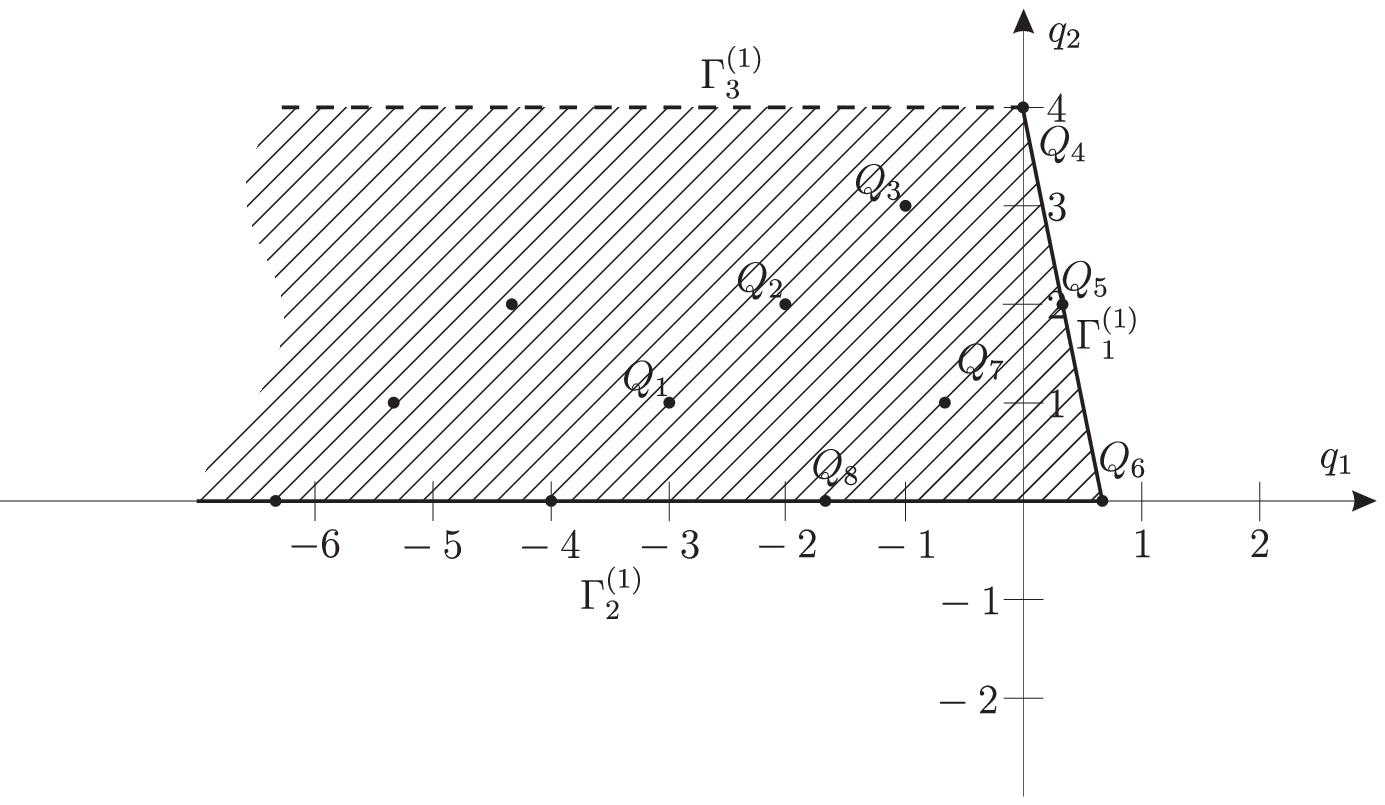,width=120mm}}
 \caption{}
\end{figure}

The periphery of the strip contains edges $\Gamma^{(1)}_j\,\,(j=1,2,3)$ with normal
vectors $N_1=(6,1),\,\, N_2=(0,-1),\,\, N_3=(0,1)$. It should take up edge
$\Gamma^{(1)}_1$ only . This edge is corresponded by the reduced
equation 
\begin{equation}
\label{6.15}h_1^{(1)}(z,\zeta) \stackrel{def}{=} \zeta^4 +18\varphi^{(l)} \zeta^2
+72{\varphi^{(l)}}^2=0
\end{equation}
Wherefrom we have
\begin{equation}
\label{6.16}\zeta^2 =-3\left(3+(-1)^{m}\right) \varphi^{(l)},\,\,\, m=1,2
\end{equation}
We obtain twelve solutions of equation  \eqref{6.15}
\begin{equation}\begin{gathered}
\label{6.17}\zeta^{(l,m,k)} =g_{1/6}^{(l,m,k)}z^{1/6},\,\,\, l=1,2,3;\,\, m,k=1,2
\end{gathered}\end{equation} where
\begin{equation}\begin{gathered}
\label{6.18,19,20}g_{1/6}^{(l,m,k)}=(-1)^{k}\sqrt{-3\left(3+(-1)^m\right)\,c_{1/3}^{(l)}}\, ,\\
\,\,\,l=1,2,3; \,\,\,m,k=1,2
\end{gathered}\end{equation}

The reduced equation is algebraic one, so it has no critical numbers. Let us compute
the carrier of the expansion for solution of equation \eqref{6.13}. The shifted carrier
of equation \eqref{6.13} is contained in a lattice, generated by vectors
$B_1=\left(\frac73,0\right),\,\, B_2=(1,1)$. The shifted carrier of solutions
\eqref{6.17} gives rise to vector $B_3=\left(-\frac16,1 \right)$. The difference
$B_2-B_3=\left(\frac76,0\right)=\frac12 B_1 \stackrel{def}{=} B_4$. Therefor, vectors
$B_1,B_2 $ and $B_3$ generate the same lattice as vectors $B_2,B_4$. Points of this
lattice can be written as
\begin{equation*}
Q=(q_1,q_2)=k(1,1) +m\left(\frac76,0\right)=\left(k+\frac{7m}6,k\right)
\end{equation*}
At the line $q_2=-1$ we have $k=-1$, and so $q_1=-1+\frac{7m}6$. As long as the cone of
the problem here is $\mathcal{K}=\left\{k<\frac16\right\}$, then the set of the
carriers of expansions $\mathbf{K}$ is
\begin{equation}
\label{6.21}\mathbf{K}=\left\{\frac{1-7n}6,n\in\mathbb{N}\right\}
\end{equation}
The expansion for solution of equation  \eqref{6.13} takes the form
\begin{equation}\begin{gathered}
\label{6.22.1}\zeta^{(l,m,k)}=g^{(l,m,k)}_{1/6} z^{1/6}
+\sum_{n}g^{(l,m,k)}_{(1-7n)/6}\,z^{(1-7n)/6},\,\,\,\\
 l=1,2,3;\,\,\quad\,
m=1,2;\,\,\quad\, k=1,2 \end{gathered}\end{equation} Coefficients $g_{1/6}^{(l,m,k)}$
are determined by expression \eqref{6.18,19,20}. Coefficient $g^{(l,m,k)}_{-1}$ takes
on a value
\begin{equation}
\label{6.22a}
g_{-1}^{(l,m,k)}=-\frac{1}{4}\\
\end{equation}
The expansion of solution with taking into account four terms takes the form
\begin{equation}
\begin{gathered}
\label{6.22b}\zeta^{(l,m,k)}=g_{{1/6}} \,{z}^{1/6}-\frac{1}{4}\,{z}^{-1}-{\frac
{7}{288}}\, \frac{\left( 17\,{g_{ {1/6}}}^{2}+63\,c_{{1/3}} \right)}{ g_{{1/6}} \left(
{g_{{1/6}}}^ {2}+9\,c_{{1/3}} \right) }\,{z}^{-13/6}-\\-{\frac {49}{1728} }\,{\frac
{17\,{g_{{1/6}}}^{4}+36\,c_{{1/3}}{g_{{1/6}}}^{2}+567\,{c_{{
1/3}}}^{2}}{{g_{{1/6}}}^{2} \left( {g_{{1/6}}}^{2}+9\,c_{{1/3}}
 \right) ^{2}}\,{z}^{-10/3}}
+ \ldots
\end{gathered}
\end{equation}

In view of \eqref{2.63} we can find  additions $u^{(l,m,k)}(z)$. We have
\begin{equation*}
\label{6.22_2}u^{(l,m,k)}(z) = C \exp \int \zeta^{(l,m,k)}(z) dz
\end{equation*}

Wherefrom we get
\begin{equation}
\begin{gathered}
\label{6.23}u^{(l,m,k)}(z)=C_1\,z^{-1/4}\, \exp \left[\frac67\,
g^{(l,m,k)}_{1/6}\,z^{7/6} + \sum^{\infty}_{n=2} \frac{6}{7(1-n)}
g^{(l,m,k)}_{(1-7n)/6}
z^{7(1-n)/6}\right]\\
\,\, l=1,2,3;\,\quad\,m=1,2;\,\quad\, k=1,2
\end{gathered}
\end{equation}
Here $C_1$ and farther $C_2$ and  $C_3$ are the arbitrary constants. Addition
$u^{(l,m,k)}(z)$ near $z\rightarrow \infty$ is the exponentially small one in those
sectors of complex plane $z$, where
\begin{equation}
\label{6.24}Re \left[g^{(l,m,k)}_{7/6}\, z^{1/6}\right]<0
\end{equation}
Thus for three expansions $G^{(1)}_{3}l$ we get four one-parameter family of additions
$G_3^{(1)}l G^1_1 mk$, where $m=1,2$ and $k=1,2$.

\section{\!\!\!\!\!\!.\,\, Exponential additions of the second level.}

Let us find exponential additions of the second level $v^{(p)}$, i.e. the additions to
solutions $u^{(l,m,k)}(z)$. The reduced equation for addition $v^{(p)}$ is
\begin{equation}
\label{6.27}M_{p}^{(2)} (z) v^{(p)}=0
\end{equation}
where operator $M_{p}^{(2)}$ is the first variation of \eqref{6.13}. Equation
\eqref{6.27} for $v=v^{(p)}$ takes the form
\begin{equation}
\begin{gathered}
\label{6.29}\frac{d^3v}{dz^3} + 4\zeta_{zz} v+ 4\zeta v_{zz} + 6\zeta_z v_z + 12
\zeta\zeta_z v+ \\
+6\zeta^2 v_z + 4\zeta ^3 v +18 \varphi ^{(l)} v_z +36\varphi ^{(l)} \zeta v
+18\varphi_z^{(l)} v=0
\end{gathered}
\end{equation}
Assumed that
\begin{equation}
\label{6.30}\frac{d \ln v}{dz}=\xi
\end{equation}
we have
\begin{equation}
\label{6.31}\frac{dv}{dz}=\xi v,\,\,\quad\, \frac{d^2v}{dz^2} =\xi_z v+\xi^2
v,\,\,\quad\, \frac{d^3 v}{dz^3}=\xi_{zz} v + 3\xi\xi_{z} v+\zeta^3 v
\end{equation}
From \eqref{6.29} we get equation
\begin{equation}\begin{gathered}
\label{6.32}\xi_{zz} + 3\, \xi\,\xi_{z} + \xi^3 + 4\zeta_{zz}
+4\xi_z\,\zeta+4\xi^2\,\zeta+ 6\,\xi\,\zeta_z +12 \zeta\zeta_z +6\,\xi\,\zeta^2 +
\\+4\,\zeta^3 +18\,\varphi^{(l)}\xi +36\,\zeta\, \varphi^{(l)}
+18\,\varphi_z^{(l)}=0
\end{gathered}\end{equation}
Monomials of equation \eqref{6.32} are corresponded by the points
\begin{equation}
\begin{gathered}
\label{6.33} M_{0,k}=\left(\frac12-\frac{7}{6}k,\,0\right),\quad
M_{1,k}=\left(\frac13-\frac{7}{6}k,1\right),\\
M_{2,k}=\left(\frac16-\frac{7}{6}k,\,2\right),\quad M_3=(0,3)\\
k=0,1,2,\ldots
\end{gathered}
\end{equation}

The carrier of the equation  \eqref{6.32} is determined by points
of the set \eqref{6.33}. The convex set forms the strip, which is
represented at fig. 4. It should examine edge $\Gamma_1^{(1)}$,
which is passing through points
\begin{equation}
\begin{gathered}
\label{6.34} Q_0=\left(\frac12,0\right),\quad Q_1=\left(\frac13,1\right),\quad
Q_2=\left(\frac16,2\right),\quad Q_3=\left(0,3\right)
\end{gathered}
\end{equation}

\begin{figure}[h!] 
 \centerline{\epsfig{file=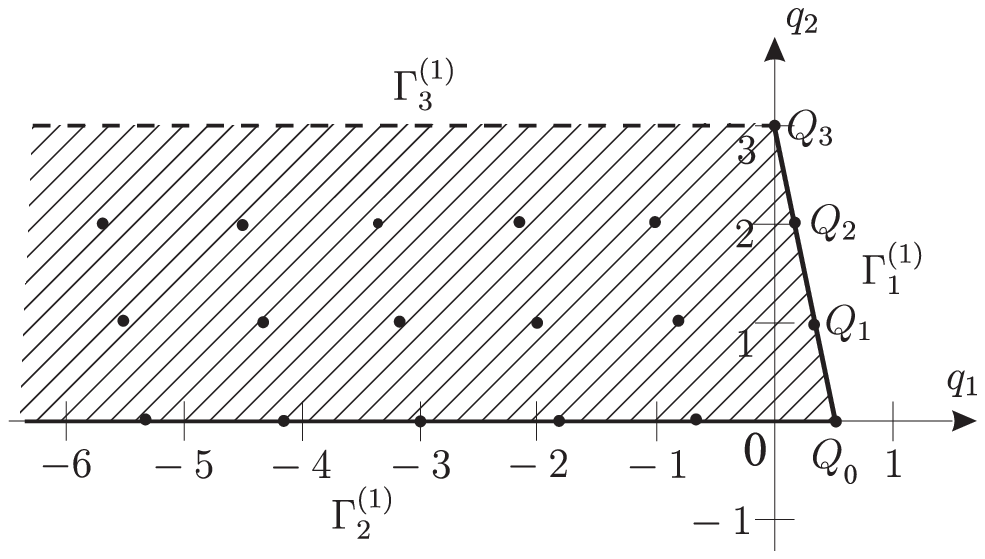,width=120mm}}
 \caption{}
\end{figure}

The reduced equation, corresponding to this edge, is
\begin{equation}
\begin{gathered}
\label{6.35}\xi^3 +4\,\xi^{2}\,\zeta
+6\,\xi\,\zeta^2+4\,\zeta^3+18\,\xi\,\varphi^{(l)}+36\,\zeta\,\varphi^{(l)} =0
\end{gathered}
\end{equation}
The basis of the lattice, corresponding to the carrier of equation
\eqref{6.32} is
\begin{equation*}
\begin{gathered}
\label{6.36}B_1=(1,1),\,\,\,\, B_2=\left(\frac76,0\right)
\end{gathered}
\end{equation*}
The solution of equation \eqref{6.35} takes the form
\begin{equation}
\begin{gathered}
\label{6.37}\xi^{(l,m,k,p)}=r_{1/6}^{(l,m,k,p)}\, z^{1/6},\,\,\,\,
m,k=1,2;\,\,\,\, l=1,2,3;\,\,\,\, p=1,2,3
\end{gathered}
\end{equation}
where $r=r_{1/6}^{(l,m,k,p)},\,\,\,p=1,2,3$  are the roots of the
equation
\begin{equation}
\begin{gathered}
\label{6.38}r^3 +4\,r^2\,g_{1/6}^{(l,m,k)}+
\left(6\,{g_{1/6}^{(l,m,k)}}^2 +18\,c_{1/3}^{(l)}\right)\,r+4\, {g_{1/6}^{(l,m,k)}}^{3}+\\
+36\,g_{1/6}^{(l,m,k)}\,c_{1/3}^{(l)}=0
\end{gathered}
\end{equation}
Equation \eqref{6.38} has the roots
\begin{equation}
\begin{gathered}
\label{6.38a}r_{1/6}^{(l,m,k,1)}=-2\,g_{1/6}^{(l,m,k)},\,\quad\,
r_{1/6}^{(l,m,k,2)}=-g_{1/6}^{(l,m,k)}+\left(-18\,c_{1/3}^{(l)}-{g_{1/6}^{(l,m,k)}}^2\right)^{1/2}\\
r_{1/6}^{(l,m,k,3)}=-g_{1/6}^{(l,m,k)}-\left(-18\,c_{1/3}^{(l)}-{g_{1/6}^{(l,m,k)}}^2\right)^{1/2}
\end{gathered}
\end{equation}
The set of carriers of expansions for solution $\mathbf{K}$
coincides with \eqref{6.21}. The expansion of solution for
$\xi^{(l,m,k,p)}$ takes the form
\begin{equation}
\begin{gathered}
\label{6.39}\xi^{(l,m,k,p)}=r_{1/6}^{(l,m,k,p)} z^{1/6}
+\sum_{n=1}^{\infty}r^{(l,m,k,p)}_{(1-7n)/6}\,z^{(1-7n)/6},\\
\,\quad\,l=1,2,3;\,\,\quad\, m=1,2;\,\,\quad\,
k=1,2;\,\quad\,p=1,2,3
\end{gathered}
\end{equation}
The computing the coefficient  $r_{-1}^{(l,m,k,p)}$ gives a result
$r_{-1}^{(l,m,k,p)}=1/6$. The expansion of solution with taking
into account three terms is
\begin{equation}
\begin{gathered}
\label{6.39a}\xi^{(l,m,k,p)}=r_{{1/6}}{z^{1/6}}+\frac16\,{z}^{-1}+ \frac {7}{72}
\,\left(34\,{g_{{1/6}}}^{3}r_{{1/6}}+{}\right.\\\left.{}+36\,c_{{1/3}}r_{
{1/6}}g_{{1/6}}+36\,c_{{1/3}}{g_{{1/6}}}^{2}+63\,c_{{1/3}}{r_{{1/6}}}^
{2}+17\,{g_{{1/6}}}^{2}{r_{{1/6}}}^{2}+{}\right.\\\left.{}+567\,{c_{{1/3}}}^{2}+17\,{g_{{1
/6}}}^{4}\right) \left(g_{1/6}\right)^{-1} \left( {g_{{1/6}}}^{2}+9\,c_{{1/3}}
\right)^{-1}\\
 \left( 8\,g_{{1/6}}r_{{1/6}}+3\,{r_{{1/6}}}^{2}+6\,{g_{{1/6}}}^{2}+18
\,c_{{1/3}} \right)^{-1} \, {z}^{-13/6}+\ldots
\end{gathered}
\end{equation}
The exponential additions $v^{(l,m,k,p)}(z)$ to solutions
$u^{(l,m,k)}(z)$ are
\begin{equation}
\begin{gathered}
\label{6.40}v^{(l,m,k,p)}(z)=C_2\,z^{1/6} \exp \left[\frac67\,
r^{(l,m,k,p)}_{1/6}\,z^{7/6} + \sum^{\infty}_{n=2} \frac{6}{7(1-n)}
r^{(l,m,k,p)}_{(1-7n)/6}
z^{7(1-n)/6}\right],\\
l=1,2,3;\,\quad\, m=1,2;\,\quad\, k=1,2;\,\quad\,p=1,2,3
\end{gathered}
\end{equation}

\section{\!\!\!\!\!\!.\,\, Exponential additions of the third level.}

Let us compute the exponential additions of the third level
$y^{(s)}$, i.e. the additions to the solutions $v^{(l,m,k,p)}(z)$.
The reduced equation for addition $y^{(s)}$ is
\begin{equation}
\label{6.41}M_{s}^{(3)} (z) y^{(s)}=0
\end{equation}
Operator $M_{s}^{(3)}$ is the first variation of \eqref{6.32}.
Equation \eqref{6.41} for $y=y^{(l,m,k,p,s)}$ takes the form
\begin{equation}
\begin{gathered}
\label{6.42}y_{zz} + 3\xi_{z} y+ 3\xi y_{z} +
3\xi^2\,y+4\,\zeta\,y_z+8\,\xi\,\zeta\,y+\\
+6\zeta_z\,y+6\zeta^2\,y+18\,\varphi^{(l)}\,y=0
\end{gathered}
\end{equation}
Using the substitute
\begin{equation}
\label{6.43}\frac{d \ln y}{dz}=\eta
\end{equation}
we obtain
\begin{equation}
\label{6.44}\frac{dy}{dz}=\eta y,\,\,\quad\, \frac{d^2y}{dz^2}
=\eta_z y+\eta^2 y
\end{equation}
From \eqref{6.44} we have equation
\begin{equation}\begin{gathered}
\label{6.45}\eta_{z} +\eta^2+ 3
\xi_z+3\xi\eta+3\xi^2+4\,\eta\,\zeta+8\,\xi\,\zeta+6\zeta_z+6\zeta^2+18\,\varphi^{(l)}=0
\end{gathered}\end{equation}
Monomials of equation \eqref{6.45} are corresponded by points
\begin{equation}
\begin{gathered}
\label{6.46}M_{0,k}=\left(\frac{1}{3}-\frac{7}{6} k,\,0\right),\quad
M_{1,k}=\left(\frac{1}{6}-\frac{7}{6}k,\,1\right),\quad M_2=\left(0,\,2\right),\\
k=0,1,2\ldots
\end{gathered}
\end{equation}
The carrier of equation \eqref{6.45} is formed by points
\eqref{6.46}. The convex set forms the strip, which is represented
at fig. 5. It should examine edge $\Gamma_1^{(1)}$, which is
passing through points
\begin{equation}
\begin{gathered}
\label{6.47}Q_0=\left(\frac13,0\right),\,\,\, Q _1=\left(\frac16,1\right),\,\,\,
Q_2=\left(0,2\right)
\end{gathered}
\end{equation}

\begin{figure}[h!] 
 \centerline{\epsfig{file=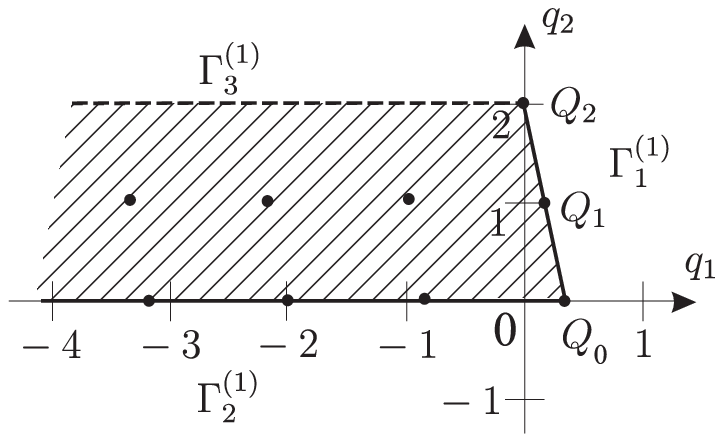,width=120mm}}
 \caption{}
\end{figure}

The reduced equation, corresponding to this edge, is
\begin{equation}
\begin{gathered}
\label{6.48}\eta^2 +
3\xi\eta+3\xi^2+4\,\eta\,\zeta+8\,\xi\,\zeta+6\,\zeta^{2}+18\,\varphi^{(l)}=0
\end{gathered}
\end{equation}

The solutions of equation \eqref{6.48} takes the form
\begin{equation}
\begin{gathered}
\label{6.50}\eta^{(l,m,k,p,s)}=q^{(l,m,k,p,s)}\, z^{1/6}\\
 l=1,2,3;\,\,\,\, m,k=1,2; \,\,\,\,p=1,2,3;\,\,\,\,s=1,2;
\end{gathered}
\end{equation}
where $q^{(l,m,k,p,s)}=q$ are the roots of equation
\begin{equation}
\begin{gathered}
\label{6.51}q^2 +\left(3\,
r_{1/6}^{(l,m,k,p)}+4\,g_{1/6}^{(l,m,k)}\right)q+8\,r_{1/6}^{(l,m,k,p)}\,g_{1/6}^{(l,m,k)}+\\+
6\,{g_{1/6}^{(l,m,k)}}^2+3\,{r_{1/6}^{(l,m,k,p)}}^2+18\,c_{1/3}^{(l)}=0
\end{gathered}
\end{equation}
The roots of equation \eqref{6.51} are
\begin{equation}
\begin{gathered}
\label{6.51a}
q_{1/6}^{(l,m,k,p,s)}=-\frac32\,r_{1/6}^{(l,m,k,p)}-2\,g_{1/6}^{(l,m,k)}+\\
+(-1)^{s-1}\,\left({-\frac34\,{r_{1/6}^{(l,m,k,p)}}^{2}
-2\,{r_{1/6}^{(l,m,k,p)}}\,g_{1/6}^{(l,m,k)}-2\,
{g_{1/6}^{(l,m,k)}}^{2}-18\,c_{1/3}^{(l)}}\right)^{1/2},\\
l=1,2,3;\,\,\,\,m,k=1,2;\,\,\,\,p=1,2,3;\,\,\,\,s=1,2;
\end{gathered}
\end{equation}
The basis of the lattice, corresponding to the carrier of equation \eqref{6.47}, is
\begin{equation*}
\begin{gathered}
\label{6.49}B_1=(1,1),\,\,\,\, B_2=\left(\frac76,0\right)
\end{gathered}
\end{equation*}
The set of carriers of expansions for  solution $\mathbf{K}$
coincides with \eqref{6.21}. The expansion of solution for
$\eta^{(l,m,k,p,s)}$ takes the form
\begin{equation}
\begin{gathered}
\label{6.52}\eta^{(l,m,k,p,s)}=q^{(l,m,k,p,s)}_{1/6} z^{1/6}
+\sum_{n=1}^{\infty}q^{(l,m,k,p,s)}_{(1-7n)/6}\,z^{(1-7n)/6},\,\,\,\\
l=1,2,3;\,\,\quad\, m=1,2;\,\,\quad\,
k=1,2;\,\,\quad\,p=1,2,3;;\,\,\quad\,s=1,2;
\end{gathered}
\end{equation}
Coefficients $q^{(l,m,k,p,s)}_{1/6},\,\,s=1,2$ are determined by formulas
\eqref{6.51a}. The computing of the coefficient $q^{(l,m,k,p,s)}_{-1}$ gives a result
$q^{(l,m,k,p,s)}_{-1}=1/6$. Exponential additions $y^{(s,p,l,m,k)}(z)$ to the solutions
$v^{(l,m,k,p)}(z)$ are
\begin{equation}
\begin{gathered}
\label{6.53}y^{(l,m,k,p,s)}(z)=C_3\,z^{1/6}\\
\exp \left[\frac67\, q^{(l,m,k,p,s)}_{1/6}\,z^{7/6} +
\sum^{\infty}_{n=2} \frac{6}{7(1-n)} q^{(l,m,k,p,s)}_{(1-7n)/6}
z^{7(1-n)/6}\right]\\
l=1,2,3;\,\quad\, m=1,2;\,\quad\,
k=1,2;\,\quad\,p=1,2,3;\,\quad\,s=1,2
\end{gathered}
\end{equation}

Thus we find three levels of the exponential additions to the expansions for solutions of
equation near point $z=\infty$.

Solution $w(z)$ at $z\rightarrow\infty$ with taking into account the exponential
additions has the expansion
\begin{equation}
\begin{gathered}
\label{6.54}w(z)=c^{(l)}_{1/3}
z^{1/3} +\frac{1}{24}\,z^{-2}+ \sum^{\infty}_{n=2} c^{(l)}_{(1-7n)/3}\, z^{{(1-7n)}/{3}}+\\
+C_1\,z^{-1/4}\,
\exp\{F_1(z)+C_2\,z^{1/6}\exp\{F_2(z)+C_3\,z^{1/6}\exp\{F_3(z)\}\}\}
\end{gathered}
\end{equation}
where $c_{1/3}^{(l)}$ can be computed  by formulas \eqref{1.61},
\eqref{1.62} and \eqref{1.63}; $F_1(z)=F_1^{(l,m,k)}(z)$,
$F_2(z)=F_2^{(l,m,k,p)}(z)$ and $F_3(z)=F_3^{(l,m,k,p,s)}(z)$,
($l=1,2,3;\,\,\,m,k=1,2;\,\,\,p=1,2,3;\,\,\,s=1,2$) are
\begin{equation}
\begin{gathered}
\label{6.55}F_1^{(l,m,k)}(z)=\frac67\, g^{(l,m,k)}_{1/6}\,z^{7/6} +
\sum^{\infty}_{n=2} \frac{6}{7(1-n)} g^{(l,m,k)}_{(1-7n)/6}
z^{7(1-n)/6}
\end{gathered}
\end{equation}
\begin{equation}
\begin{gathered}
\label{6.56}F_2^{(p,l,m,k)}(z)=\frac67\,
r^{(l,m,k,p)}_{1/6}\,z^{7/6} + \sum^{\infty}_{n=2} \frac{6}{7(1-n)}
r^{(l,m,k,p)}_{(1-7n)/6} z^{7(1-n)/6}
\end{gathered}
\end{equation}
\begin{equation}
\begin{gathered}
\label{6.57}F_3^{(l,m,k,p,s)}(z)=\frac67\,
q^{(l,m,k,p,s)}_{1/6}\,z^{7/6} + \sum^{\infty}_{n=2}
\frac{6}{7(1-n)} q^{(l,m,k,p,s)}_{(1-7n)/6} z^{7(1-n)/6}
\end{gathered}
\end{equation}
Coefficients $g^{(l,m,k)}_{1/6}$, $r^{(l,m,k,p)}_{1/6}$ and $q^{(l,m,k,p,s)}_{1/6}$ are
defined by formulas \eqref{6.18,19,20}, \eqref{6.38a} and \eqref{6.51a}. The other
coefficients are computed sequentially.

\section{\!\!\!\!\!\!.\,\, Conclusion.}
Let us formulate the results of this work.

All the power asymptotic forms for equation \eqref{1.7} were found.

We also found all the power expansions, corresponding to these
asymptotic forms. We denote the obtained families as  $G_1^{(0)}1$,
$G_1^{(0)}2$, $G_1^{(0)}3$, $G_1^{(0)}4$, $G_1^{(1)}1$, $G_2^{(1)}1$
and $G_2^{(1)}2$ (these expansions converge for sufficiently small
$|z|$). The existence and analyticity of these expansions follow
from Cauchy theorem.

We found three families of expansions near $z=\infty$. That are
families $G_3^{(1)}l\,\,(l=1,2,3)$, described by formulas
\eqref{1.61}, \eqref{1.62} and \eqref{1.63}. For each of these
expansions we found four exponential additions
$G_3^{(1)}lG_1^1mk\,\,(m,k=1,2)$ expressed by formula \eqref{6.23}.
For them it was computed exponential additions
$G_3^{(1)}lG_1^1mkG_1^{(1)}p\,\,(m,k=1,2;p=1,2,3)$, and then for
them the proper exponential additions  $G_3^{(1)}lG_1^{(1)}mk
G_1^{(1)}pG_1^{(1)}s\,\,(m,k=1,2;p=1,2,3;s=1,2)$ were found too.

Families $G_2^{(1)}l\,\,$ and $G_2^{(1)}2$ were first found in the
paper \cite{Kudryashov0/5, Kudryashov0/6, Kudryashov10,
Pickering01}. However the structure of expansions $G_2^{(1)}1$ and
$G_2^{(2)}2$ was not discussed earlier. The other families of
expansions  of solution are found for the first time.

Comparing the power expansions of equation \eqref{1.7} with power
expansions of Painlev\'{e} equations $P_1 \div P_6$ \cite{Golubev01,
Bruno03,Bruno04,Gromak01} we note, that they differ. This fact can
be interpreted as the additional prove for the hypothesis, that the
fourth-order equation \eqref{1.7} determines new transcendental
functions just as equations $P_1 \div P_6$.

\end{document}